\documentclass[aps,prl,twocolumn,floatfix]{revtex4-1}
\usepackage{graphicx}
\usepackage{amsmath}
\usepackage{braket}
\usepackage{tikz}
\begin{document}

\title{Quench-produced solitons in a box-trapped Bose-Einstein condensate}
\begin{abstract}
We describe a protocol to prepare solitons in a quasi-1d box-trapped Bose-Einstein condensate using only a quench of the isotropic s-wave scattering length. A quench to exactly four times the initial 1d coupling strength creates one soliton at each boundary of the box, which then propagate in a uniform background density and collide with one another. No nonsolotonic excitations are created during the quench. The procedure is robust against imperfections in the scattering length ramp rate and a mismatch of the final scattering length.   
\end{abstract}

\author{E. J. Halperin}
\affiliation{JILA, NIST, and Department of Physics, University of Colorado, Boulder, Colorado 80309-0440, USA}
\author{J. L. Bohn}
\affiliation{JILA, NIST, and Department of Physics, University of Colorado, Boulder, Colorado 80309-0440, USA}
\date{\today}
\maketitle

\emph{Introduction.---}
Solitons appear in a variety of physical systems, such as shallow water, nonlinear optics, and interacting Bose-Einstein condensates (BEC)~\cite{frantzeskakis2010dark}. Solitons, or ``solitary waves'', are localized modulations  – whether of height, intensity, or density  – which propagate without changing their shape or velocity. They are a hallmark of certain nonlinear problems, often where a complete spectrum is presently unobtainable. As opposed to phonons, which treat a nonlinear system as essentially linear, solitons rely on and demonstrate the nonlinear nature of the system. In a linear quantum system, such a density modulation would naturally disperse in a uniform background, and thus solitons in a BEC represent a compromise between interatomic interactions and generic wave-packet dispersion. Solitons provide an avenue for exploring manifestly nonlinear properties of interacting Bose gasses.

As a result, a great deal of effort has been put into producing solitons in a laboratory setting. Several ingenious methods have been devised and experimentally implemented for making soltions in a quasi-1d Bose gas. Soltions were first created by directly phase imprinting in a harmonically trapped BEC via carefully controlled laser fields~\cite{burger1999dark,denschlag2000generating}. Here the condensate is phase shifted by pulsing on potential to create a phase jump which is characteristic of solitons. The imprinted phase discontinuity generates solitons as well as other structures~\cite{burger1999dark} in the gas. These solitons are localized dips in density, i.e. gray solitons. Bright solitons, which are areas of higher density, have also been formed via quenches to negative scattering lengths~\cite{strecker2003bright, carr2004pulsed}, where fluctuations in the positive scattering length gas naturally coalesce after a quench to a negative scattering length. Similar to the phase imprinting method, solitons have been formed by moving from one trap to another~\cite{weller2008experimental}, where a BEC is split into two pieces which are then allowed to collide, creating solitonic and other excitations. 

In general, these methods create excitations that are characteristic of, although often not exactly, solitons, and in turn lead to additional excitations.  Moreover, in typical experiments the solitons are created in a harmonically trapped BEC, and thus propagate against a spatially varying background density, complicating their dynamics~\cite{astrakharchik2004motion}. In this Letter we take advantage of two fairly recently developed experimental technologies – box traps~\cite{gaunt2013bose} and the ability to suddenly change the scattering length~\cite{makotyn2014universal} – to propose a relatively clean protocol for preparing solitons. In the a box trap, the resulting shape of the density profile near the walls serves as a seed for the solitons to come, which emerge upon swiftly changing the scattering length to certain predetermined values.  As an additional benefit, the resulting solitons propagate in the uniform-density interior of the box trap.

The method originates in a fascinating mathematical prediction from inverse scattering theory~\cite{gamayun2015fate,Franchini_2015}, which posits that in a quench of the coupling strength $g$ from $g \rightarrow n^2 g$, where $n$ is an integer, a soliton in an infinite uniform density BEC splits into $2 n - 1$ solitons. This theory predicts that in a quench to any value other than a perfect square multiple of the initial scattering length, the quench will create solitions along with additional excitations. A interesting special case of a soliton is a dark soliton, which is stationary and causes the density to go to zero at its center.  After a quench in a uniform gas of a dark soliton, the inverse scattering theory predicts that there will be $n - 1$ solitons moving to the left and $n - 1$ to the right, with one, now narrower, dark soliton remaining stationary in the original location. 
 
Motivated by this mathematical curiosity, this Letter proposes an operationally straightforward and robust way to create solitons in a box-trapped BEC. The ground state density of a large box-trapped BEC must go from nearly uniform in the center to zero at the edges, and does so such that the wave function looks exactly like half of a dark soliton. Upon a quench of the ground state BEC in a box from $g \rightarrow 4g$, each ``half-soliton'' at the boundary of the box generates one gray soliton, which are launched towards the center of the trap. No other excitations are created during the idealized quench; aside from the two travelling solitons and the boundary conditions, the gas presents a uniform background density. After verifying the efficacy of the procedure numerically, we discuss robustness to experimental considerations. In particular, we look at a quench to a value that is close to, but not exactly, $4g$ as well as a finite ramp time from $g \rightarrow 4g$.

\emph{Demonstration of the protocol.---}
We numerically verify the inverse scattering theory claim~\cite{gamayun2015fate,Franchini_2015} for a soliton in a uniform BEC. We simulate the Gross-Pitaevskii equation in a box with either periodic or hard wall boundary conditions, which in 1d is given by
\begin{align}
i\hbar\frac{\partial\Psi}{\partial t} = \left[-\frac{\hbar^2}{2m} \frac{\partial^2}{\partial x^2} + V + N g\vert\Psi\vert^2 \right] \Psi,
\label{GPE}
\end{align}
where $\Psi$ is the many-body order parameter of the $N$ particle Bose gas, related to the single particle wave function $\phi$ by $\Psi = \sqrt{N} \phi$. In 1d $\Psi$ describes the linear number density. We set the external potential $V(x) = 0$ and $g$ the 1d coupling constant, which is related to the 3d scattering length $a$ by~\cite{olshanii1998atomic,astrakharchik2004quasi}
\begin{align}
    g = \frac{2 \hbar^2 a}{m a_\rho^2} \frac{1}{1 - A a_\rho / a},
\end{align}
with $A \approx 1.03$, and $a_\rho$ the oscillator length in the transverse direction. Although we are considering here a harmonic trap in the transverse direction, the discussion would be much the same for arbitrary transverse confinement, as long as that confinement is sufficiently tight. For $ 0 < a \ll a_\rho$, then
\begin{align}
    g \approx \frac{2 \hbar^2 a}{m a_\rho^2},
\end{align}
so in this case a quench of the 3d scattering length by a factor of 4 corresponds to a quench of the 1d coupling strength by a factor of 4. Outside of this limit special care must be taken near confinement induced resonances.  In order to quench the 1d coupling strength by a factor of 4, the required final 3d scattering length $a_f$ in terms of the initial 3d scattering length $a$ is 
\begin{align}
    a_f = \frac{4a_ \rho a}{a_\rho + 3 A a},
    \label{eq:3d_rule}
\end{align}
which reduces to $a_f = 4a$ in the case of $a \ll a_\rho$.

We find the ground state solution of Eq.\eqref{GPE} via imaginary time evolution~\cite{chiofalo2000ground} and evolve the solution in time using a time-splitting psuedo-spectral method~\cite{bao2012mathematical}. The wave function for a single soliton in a uniform BEC has the form~\cite{pitaevskii2003bose}
\begin{align}
    \psi = \sqrt{n}\left[ i\frac{v}{c} + \sqrt{1 - \frac{v^2}{c^2}} \tanh\left( \frac{x - vt}{\sqrt{2}\xi} \sqrt{1 - \frac{v^2}{c^2}} \right) \right],
    \label{eq:soliton}
\end{align}
with $n$ the background density, $v$ the velocity, $c = \sqrt{gn/m}$ the speed of sound, and $\xi = \hbar / (\sqrt{m g n})$, is the healing length, which is sometimes defined with a factor of $\sqrt{2}$ in the denominator. The soliton is entirely determined by the single parameter $v$, which fixes both its velocity and depth. It is well known that solitons are unstable outside of the the quasi-1d regime, where they decay into vortices and other excitations.  We limit our discussion to tightly confined quasi-1d condensates, where there is no transverse instability. We thus need to be in the regime where $N a / L < 0.6$~\cite{PhysRevA.60.R2665,frantzeskakis2010dark} and $a_\rho/ L \ll 1$, in which the soliton will be stable and the gas will be quasi-1d.  Although the idealized protocol is independent of the specific parameters, we here solve the Gross-Pitaevskii equation in a $100~\mu\text{m}$ box, and a healing length of $\xi = 3.1~\mu\text{m}$. This can be achieved for example with $5000$ $^{39} K$ atoms at $a_{3d} = 20~a_0$ with $a_\rho = 1 \mu\text{m}$. Here $N a / L = 0.05$ and $a_\rho / L = 0.005$. The speed of sounds $c = 0.5~\text{mm/s}$, and the final 3d scattering length that gives four times the initial 1d coupling strength is $a_f = 79.7~a_0$, as found via Eq.~\eqref{eq:3d_rule}.

\begin{figure}[bt]
  \centering
  \includegraphics[width = 0.48\textwidth]{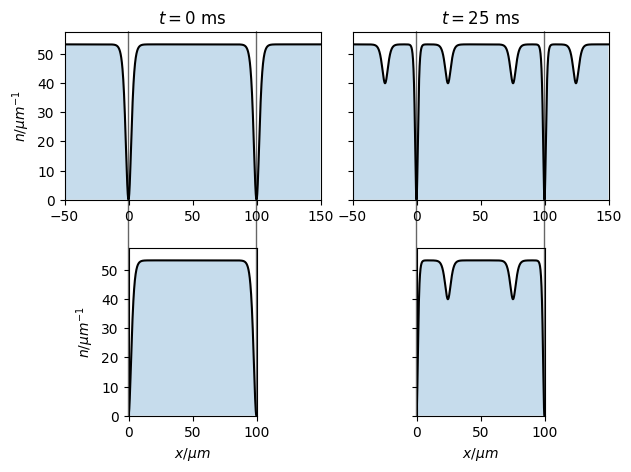}
  \caption{Comparing the dynamics of two dark solitons after a quench in a uniform BEC (top) to the ground state in a box trap following the same quench (bottom), shown before (left) and $25~\text{ms}$ after (right) a quench to four times the initial coupling strength, with parameters as described in the text. The density in the box trap exactly matches that in the uniform gas both before and after the quench within the boundaries of the box. Gray lines emphasize this correspondence.}
  \label{demonstration}
\end{figure}

Figure~\ref{demonstration} (top left) shows two dark solitons in uniform BEC well-separated by a distance $L = 100~\mu\text{m}$, with one at $x = 0$ and one at $x = L$, simulated using periodic boundary conditions.  The solitons are far enough apart that they are essentially independent and are well described by Eq.~\eqref{eq:soliton}. Each soliton splits into three solitons following a quench from $g \rightarrow 4g$. This is shown at a later time (top right). Inverse scattering theory gives quantitative predictions for the shape and velocity, with which our numerics agree to one part in $10^{4}$. The newly created solitons have a minimum density that is $3/4$ of the background density and move at $\sqrt{3}/2$ of the final speed of sound, as theoretically predicted~\cite{gamayun2015fate}. It is important to make a distinction here between the effective background density and the average density, which are not the same in a box. The solitons propagate in a background density that is slightly higher than the average density, due to the finite nature of the box. This can be seen in Fig.~\ref{demonstration}, as the average density is $50~\mu\text{m}^{-1}$ but the background density is seen to be $53~\mu\text{m}^{-1}$.

On the bottom row of Fig.~\ref{demonstration}, we show simulated density profiles for a box-trapped BEC, before (bottom left) and $25~\text{ms}$ after (bottom right) a quench from $g \rightarrow 4g$. One can see how the density profile of the two solitons in the uniform gas exactly matches that of the box-trapped BEC, both before and after the quench. Furthermore, the narrower dark solitons, without the newly created gray solitons, then match the ground state of the  more strongly interacting gas near the boundaries of the box. The expelled volume does not change during the quench and thus the background density remains the same. From Fig.~\ref{demonstration}, the correspondence between the two scenarios is evident. The dynamics of the solitons in the uniform density BEC preserves the boundary conditions of the box at all times. Thus, using a ``method of images'' we can exactly map the problem of two dark solitons separated by distance $L$ in a \emph{uniform} BEC to the ground state of a \emph{trapped} BEC. In this one way may produce solitons in a box by a isotropic quench of the 1d coupling strength.

\emph{Robustness to experimental considerations.---}
The simulation shown in Fig.~\ref{demonstration} is idealized because it assumes an instantaneous quench from $g$ to exactly $4g$. Given that the protocol originates in a mathematical theory~\cite{gamayun2015fate,Franchini_2015}, it is fair to wonder whether the perfect solitons are badly marred if these conditions are not met.  In this section we verify that this is not so: the solitons are strikingly robust to imperfections in their production.   

Within the 1d regime, we look at errors produced by noninstantaneous quench times and mismatch in the final coupling strength. There are two main types of error produced here. First the solitons may not be the correct width or depth; second there are density modulations in the bulk due to the production of phonon modes. For each of these quenches, we fit each gray soliton to a function of the form
\begin{align}
    \psi(x;d,w) = \sqrt{n}\left[ i \sqrt{1 - d} + \sqrt{d} \tanh\left(\sqrt{\frac{3}{8}} \frac{x - v t}{ \xi_f w}  \right) \right],
\end{align}
with $\xi_f$ the final healing length. As opposed to Eq.\eqref{eq:soliton}, the term inside the hyperbolic tangent is independent of d in order to decouple variations in width and depth. The ideal soliton has $w~=~1$ and $d~=~0.25$. We quantify the error in width $\Delta w$ and height $\Delta d$ simply as the fractional error from the predicted values. So $\Delta w~=~w - 1$ and $\Delta d~=~4(d - 0.25)$. When $\Delta d$ is positive (negative), the soliton is deeper (shallower) than expected, and when $\Delta w$ is positive (negative), the soliton is wider (narrower) than expected. 

\begin{figure}[tb]
  \centering
  \includegraphics[width = 0.48\textwidth]{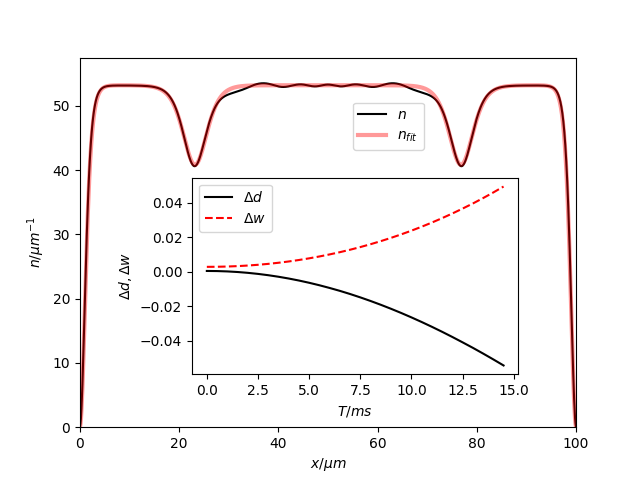}
  \caption{The density profile $25~\text{ms}$ after quench from $g \rightarrow 4.2 g$ in a $100~\mu\text{m}$ box with a quench time of $4~\text{ms}$. The inset shows the variations in the expected soliton height and width as a function of ramp time, assuming an ideal final scattering length. As ramp time is increased, the solitons become wider and shallower.}
  \label{robust}
\end{figure}

Figure~\ref{robust} shows an example of the density profile following an imperfect quench (solid black line), along with the the best fit function we use to calculate $\Delta w$ and $\Delta d$ (shaded red line). The final coupling is $5 \%$ off from the ideal value, i.e. $g_f = 4.2 g$, while the quench is not instantaneous, and instead takes $4~\text{ms}$. The main excitations are well described by something solitonic, however with different depth and width than one would expect. The solitons are slightly shallower than predicted here, as well as wider. Although a higher scattering length would lead to narrower solitons, the finite quench time offsets this, as the solitons are produced during the quench, when the scattering length is less than its final value. In addition, there are phonon modes which herald the arrival of the solitons. These phonons originate at the edges of the box and propagate faster than the solitons at the speed of sound, so they appear in the center of the box before the solitons. As time goes on for the scenario shown in Fig.~\ref{robust}, the solitons narrow and deepen, and so the error, which is initially dominated by the mismatch in soliton shape, will transfer into nonlocalized phonon modes. 

The inset of Fig.~\ref{robust} shows the deviations from the predicted width (dashed red line) and depth (solid black line) of the solitons. In this case the quench is not instantaneous however the final coupling strength is exactly $4g$. This shows that at longer quench times the produced solitons are wider and shallower. We evaluate the fractional error in width and depth after $25~\text{ms}$, when the solitons are well-separated from each other and the boundaries. 

\begin{figure}[tb]
  \centering
  \includegraphics[width = 0.48\textwidth]{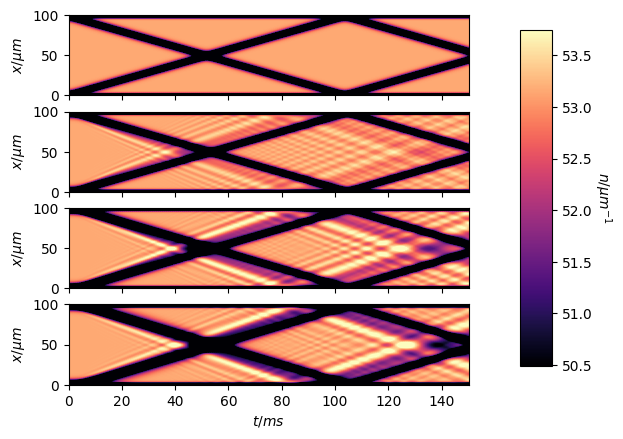}
  \caption{Density map showing the time dynamics for quenches of $0,4,8,12~\text{ms}$ top to bottom, showing the solitons and other structures created during these quenches. For the instantaneous quench, only solitons are created, which propagate undisturbed. For slower quenches, some other modes are populated, however the dynamics are still dominated by solitons.}
  \label{time_quench}
\end{figure}

It is interesting to consider the complete dynamics of the solitons during a finite quench. Figure~\ref{time_quench} shows the density as a function of time for four different quench durations. Here we show quench times of $0,4,8$ and $12~\text{ms}$, where in each case the quench is to $4g$. The top panel shows the idealized quench, where the solitons are created at the boundary and propagate through the bulk, before colliding with one another around $50~\text{ms}$. The solitons experience a slight phase shift during the collision~\cite{akhmediev1993first}, but emerge with the same velocity and shape. Near the boundaries of the box, the solitons slow down and reverse directions, essentially moving in a potential proportional to the background density~\cite{astrakharchik2004motion}. The second panel shows a $4~\text{ms}$ quench, where there is a slight delay in when the solitons move away from the boundaries. Many phonons are created here, although the solitons themselves are virtually unchanged. The phonons begin as local excitations which move faster than solitons and eventually spread out over the bulk of the gas, as can be seen by the striations in the second panel. These modes interfere with one another as they move through the gas. In the third panel, which shows an $8~\text{ms}$ quench, one can start to see the solitons widen at early times, and the phonon modes become larger in amplitude. The fourth panel is the longest quench of $12~\text{ms}$, where the previously described effects are amplified. Here one can see the solitons narrow over time. Despite this long quench length, the dynamics are still dominated by solotonic features, demonstrating the robustness of the procedure. 

\emph{Conclusion.---}
We have proposed and numerically verified a protocol to create solitons in a box using a quench of the coupling strength to four times its initial value. This protocol is seen to be robust to two possible sources of error. This protocol could also create solitons at very large scattering lengths where the ground state could not be adiabatically prepared due to three-body loss, as the ground state could be prepared at comparatively low scattering length prior to a quench. Soliton trains could also be formed by quenching to another perfect square of the initial coupling strength, for example a quench from $g \rightarrow 9 g$ would create two gray solitons at each boundary. 
\\ \\
This material is based upon work supported by the National Science Foundation under Grant Number PHY 1734006 and Grant Number PHY 1806971.
\bibliographystyle{apsrev4-1} 
\bibliography{box} 

\begin{thebibliography}{18}%
\makeatletter
\providecommand \@ifxundefined [1]{%
 \@ifx{#1\undefined}
}%
\providecommand \@ifnum [1]{%
 \ifnum #1\expandafter \@firstoftwo
 \else \expandafter \@secondoftwo
 \fi
}%
\providecommand \@ifx [1]{%
 \ifx #1\expandafter \@firstoftwo
 \else \expandafter \@secondoftwo
 \fi
}%
\providecommand \natexlab [1]{#1}%
\providecommand \enquote  [1]{``#1''}%
\providecommand \bibnamefont  [1]{#1}%
\providecommand \bibfnamefont [1]{#1}%
\providecommand \citenamefont [1]{#1}%
\providecommand \href@noop [0]{\@secondoftwo}%
\providecommand \href [0]{\begingroup \@sanitize@url \@href}%
\providecommand \@href[1]{\@@startlink{#1}\@@href}%
\providecommand \@@href[1]{\endgroup#1\@@endlink}%
\providecommand \@sanitize@url [0]{\catcode `\\12\catcode `\$12\catcode
  `\&12\catcode `\#12\catcode `\^12\catcode `\_12\catcode `\%12\relax}%
\providecommand \@@startlink[1]{}%
\providecommand \@@endlink[0]{}%
\providecommand \url  [0]{\begingroup\@sanitize@url \@url }%
\providecommand \@url [1]{\endgroup\@href {#1}{\urlprefix }}%
\providecommand \urlprefix  [0]{URL }%
\providecommand \Eprint [0]{\href }%
\providecommand \doibase [0]{http://dx.doi.org/}%
\providecommand \selectlanguage [0]{\@gobble}%
\providecommand \bibinfo  [0]{\@secondoftwo}%
\providecommand \bibfield  [0]{\@secondoftwo}%
\providecommand \translation [1]{[#1]}%
\providecommand \BibitemOpen [0]{}%
\providecommand \bibitemStop [0]{}%
\providecommand \bibitemNoStop [0]{.\EOS\space}%
\providecommand \EOS [0]{\spacefactor3000\relax}%
\providecommand \BibitemShut  [1]{\csname bibitem#1\endcsname}%
\let\auto@bib@innerbib\@empty
\bibitem [{\citenamefont {Frantzeskakis}(2010)}]{frantzeskakis2010dark}%
  \BibitemOpen
  \bibfield  {author} {\bibinfo {author} {\bibfnamefont {D.}~\bibnamefont
  {Frantzeskakis}},\ }\href@noop {} {\bibfield  {journal} {\bibinfo  {journal}
  {J Phys A - Math Theor}\ }\textbf {\bibinfo {volume} {43}},\ \bibinfo {pages}
  {213001} (\bibinfo {year} {2010})}\BibitemShut {NoStop}%
\bibitem [{\citenamefont {Burger}\ \emph {et~al.}(1999)\citenamefont {Burger},
  \citenamefont {Bongs}, \citenamefont {Dettmer}, \citenamefont {Ertmer},
  \citenamefont {Sengstock}, \citenamefont {Sanpera}, \citenamefont
  {Shlyapnikov},\ and\ \citenamefont {Lewenstein}}]{burger1999dark}%
  \BibitemOpen
  \bibfield  {author} {\bibinfo {author} {\bibfnamefont {S.}~\bibnamefont
  {Burger}}, \bibinfo {author} {\bibfnamefont {K.}~\bibnamefont {Bongs}},
  \bibinfo {author} {\bibfnamefont {S.}~\bibnamefont {Dettmer}}, \bibinfo
  {author} {\bibfnamefont {W.}~\bibnamefont {Ertmer}}, \bibinfo {author}
  {\bibfnamefont {K.}~\bibnamefont {Sengstock}}, \bibinfo {author}
  {\bibfnamefont {A.}~\bibnamefont {Sanpera}}, \bibinfo {author} {\bibfnamefont
  {G.~V.}\ \bibnamefont {Shlyapnikov}}, \ and\ \bibinfo {author} {\bibfnamefont
  {M.}~\bibnamefont {Lewenstein}},\ }\href@noop {} {\bibfield  {journal}
  {\bibinfo  {journal} {Phys. Rev. Lett.}\ }\textbf {\bibinfo {volume} {83}},\
  \bibinfo {pages} {5198} (\bibinfo {year} {1999})}\BibitemShut {NoStop}%
\bibitem [{\citenamefont {Denschlag}\ \emph {et~al.}(2000)\citenamefont
  {Denschlag}, \citenamefont {Simsarian}, \citenamefont {Feder}, \citenamefont
  {Clark}, \citenamefont {Collins}, \citenamefont {Cubizolles}, \citenamefont
  {Deng}, \citenamefont {Hagley}, \citenamefont {Helmerson}, \citenamefont
  {Reinhardt} \emph {et~al.}}]{denschlag2000generating}%
  \BibitemOpen
  \bibfield  {author} {\bibinfo {author} {\bibfnamefont {J.}~\bibnamefont
  {Denschlag}}, \bibinfo {author} {\bibfnamefont {J.~E.}\ \bibnamefont
  {Simsarian}}, \bibinfo {author} {\bibfnamefont {D.~L.}\ \bibnamefont
  {Feder}}, \bibinfo {author} {\bibfnamefont {C.~W.}\ \bibnamefont {Clark}},
  \bibinfo {author} {\bibfnamefont {L.~A.}\ \bibnamefont {Collins}}, \bibinfo
  {author} {\bibfnamefont {J.}~\bibnamefont {Cubizolles}}, \bibinfo {author}
  {\bibfnamefont {L.}~\bibnamefont {Deng}}, \bibinfo {author} {\bibfnamefont
  {E.~W.}\ \bibnamefont {Hagley}}, \bibinfo {author} {\bibfnamefont
  {K.}~\bibnamefont {Helmerson}}, \bibinfo {author} {\bibfnamefont {W.~P.}\
  \bibnamefont {Reinhardt}},  \emph {et~al.},\ }\href@noop {} {\bibfield
  {journal} {\bibinfo  {journal} {Science}\ }\textbf {\bibinfo {volume}
  {287}},\ \bibinfo {pages} {97} (\bibinfo {year} {2000})}\BibitemShut
  {NoStop}%
\bibitem [{\citenamefont {Strecker}\ \emph {et~al.}(2003)\citenamefont
  {Strecker}, \citenamefont {Partridge}, \citenamefont {Truscott},\ and\
  \citenamefont {Hulet}}]{strecker2003bright}%
  \BibitemOpen
  \bibfield  {author} {\bibinfo {author} {\bibfnamefont {K.}~\bibnamefont
  {Strecker}}, \bibinfo {author} {\bibfnamefont {G.}~\bibnamefont {Partridge}},
  \bibinfo {author} {\bibfnamefont {A.}~\bibnamefont {Truscott}}, \ and\
  \bibinfo {author} {\bibfnamefont {R.~G.}\ \bibnamefont {Hulet}},\ }\href@noop
  {} {\bibfield  {journal} {\bibinfo  {journal} {New J. Phys.}\ }\textbf
  {\bibinfo {volume} {5}},\ \bibinfo {pages} {73} (\bibinfo {year}
  {2003})}\BibitemShut {NoStop}%
\bibitem [{\citenamefont {Carr}\ and\ \citenamefont
  {Brand}(2004)}]{carr2004pulsed}%
  \BibitemOpen
  \bibfield  {author} {\bibinfo {author} {\bibfnamefont {L.}~\bibnamefont
  {Carr}}\ and\ \bibinfo {author} {\bibfnamefont {J.}~\bibnamefont {Brand}},\
  }\href@noop {} {\bibfield  {journal} {\bibinfo  {journal} {Phys. Rev. A}\
  }\textbf {\bibinfo {volume} {70}},\ \bibinfo {pages} {033607} (\bibinfo
  {year} {2004})}\BibitemShut {NoStop}%
\bibitem [{\citenamefont {Weller}\ \emph {et~al.}(2008)\citenamefont {Weller},
  \citenamefont {Ronzheimer}, \citenamefont {Gross}, \citenamefont {Esteve},
  \citenamefont {Oberthaler}, \citenamefont {Frantzeskakis}, \citenamefont
  {Theocharis},\ and\ \citenamefont {Kevrekidis}}]{weller2008experimental}%
  \BibitemOpen
  \bibfield  {author} {\bibinfo {author} {\bibfnamefont {A.}~\bibnamefont
  {Weller}}, \bibinfo {author} {\bibfnamefont {J.}~\bibnamefont {Ronzheimer}},
  \bibinfo {author} {\bibfnamefont {C.}~\bibnamefont {Gross}}, \bibinfo
  {author} {\bibfnamefont {J.}~\bibnamefont {Esteve}}, \bibinfo {author}
  {\bibfnamefont {M.}~\bibnamefont {Oberthaler}}, \bibinfo {author}
  {\bibfnamefont {D.}~\bibnamefont {Frantzeskakis}}, \bibinfo {author}
  {\bibfnamefont {G.}~\bibnamefont {Theocharis}}, \ and\ \bibinfo {author}
  {\bibfnamefont {P.}~\bibnamefont {Kevrekidis}},\ }\href@noop {} {\bibfield
  {journal} {\bibinfo  {journal} {Phys. Rev. Lett.}\ }\textbf {\bibinfo
  {volume} {101}},\ \bibinfo {pages} {130401} (\bibinfo {year}
  {2008})}\BibitemShut {NoStop}%
\bibitem [{\citenamefont {Astrakharchik}\ and\ \citenamefont
  {Pitaevskii}(2004)}]{astrakharchik2004motion}%
  \BibitemOpen
  \bibfield  {author} {\bibinfo {author} {\bibfnamefont {G.}~\bibnamefont
  {Astrakharchik}}\ and\ \bibinfo {author} {\bibfnamefont {L.}~\bibnamefont
  {Pitaevskii}},\ }\href@noop {} {\bibfield  {journal} {\bibinfo  {journal}
  {Phys. Rev. A}\ }\textbf {\bibinfo {volume} {70}},\ \bibinfo {pages} {013608}
  (\bibinfo {year} {2004})}\BibitemShut {NoStop}%
\bibitem [{\citenamefont {Gaunt}\ \emph {et~al.}(2013)\citenamefont {Gaunt},
  \citenamefont {Schmidutz}, \citenamefont {Gotlibovych}, \citenamefont
  {Smith},\ and\ \citenamefont {Hadzibabic}}]{gaunt2013bose}%
  \BibitemOpen
  \bibfield  {author} {\bibinfo {author} {\bibfnamefont {A.~L.}\ \bibnamefont
  {Gaunt}}, \bibinfo {author} {\bibfnamefont {T.~F.}\ \bibnamefont
  {Schmidutz}}, \bibinfo {author} {\bibfnamefont {I.}~\bibnamefont
  {Gotlibovych}}, \bibinfo {author} {\bibfnamefont {R.~P.}\ \bibnamefont
  {Smith}}, \ and\ \bibinfo {author} {\bibfnamefont {Z.}~\bibnamefont
  {Hadzibabic}},\ }\href@noop {} {\bibfield  {journal} {\bibinfo  {journal}
  {Phy. Rev. Lett.}\ }\textbf {\bibinfo {volume} {110}},\ \bibinfo {pages}
  {200406} (\bibinfo {year} {2013})}\BibitemShut {NoStop}%
\bibitem [{\citenamefont {Makotyn}\ \emph {et~al.}(2014)\citenamefont
  {Makotyn}, \citenamefont {Klauss}, \citenamefont {Goldberger}, \citenamefont
  {Cornell},\ and\ \citenamefont {Jin}}]{makotyn2014universal}%
  \BibitemOpen
  \bibfield  {author} {\bibinfo {author} {\bibfnamefont {P.}~\bibnamefont
  {Makotyn}}, \bibinfo {author} {\bibfnamefont {C.~E.}\ \bibnamefont {Klauss}},
  \bibinfo {author} {\bibfnamefont {D.~L.}\ \bibnamefont {Goldberger}},
  \bibinfo {author} {\bibfnamefont {E.}~\bibnamefont {Cornell}}, \ and\
  \bibinfo {author} {\bibfnamefont {D.~S.}\ \bibnamefont {Jin}},\ }\href@noop
  {} {\bibfield  {journal} {\bibinfo  {journal} {Nat. Phys.}\ }\textbf
  {\bibinfo {volume} {10}},\ \bibinfo {pages} {116} (\bibinfo {year}
  {2014})}\BibitemShut {NoStop}%
\bibitem [{\citenamefont {Gamayun}\ \emph {et~al.}(2015)\citenamefont
  {Gamayun}, \citenamefont {Bezvershenko},\ and\ \citenamefont
  {Cheianov}}]{gamayun2015fate}%
  \BibitemOpen
  \bibfield  {author} {\bibinfo {author} {\bibfnamefont {O.}~\bibnamefont
  {Gamayun}}, \bibinfo {author} {\bibfnamefont {Y.~V.}\ \bibnamefont
  {Bezvershenko}}, \ and\ \bibinfo {author} {\bibfnamefont {V.}~\bibnamefont
  {Cheianov}},\ }\href@noop {} {\bibfield  {journal} {\bibinfo  {journal}
  {Phys. Rev. A}\ }\textbf {\bibinfo {volume} {91}},\ \bibinfo {pages} {031605}
  (\bibinfo {year} {2015})}\BibitemShut {NoStop}%
\bibitem [{\citenamefont {Franchini}\ \emph {et~al.}(2015)\citenamefont
  {Franchini}, \citenamefont {Gromov}, \citenamefont {Kulkarni},\ and\
  \citenamefont {Trombettoni}}]{Franchini_2015}%
  \BibitemOpen
  \bibfield  {author} {\bibinfo {author} {\bibfnamefont {F.}~\bibnamefont
  {Franchini}}, \bibinfo {author} {\bibfnamefont {A.}~\bibnamefont {Gromov}},
  \bibinfo {author} {\bibfnamefont {M.}~\bibnamefont {Kulkarni}}, \ and\
  \bibinfo {author} {\bibfnamefont {A.}~\bibnamefont {Trombettoni}},\ }\href
  {\doibase 10.1088/1751-8113/48/28/28ft01} {\bibfield  {journal} {\bibinfo
  {journal} {J Phys A - Math Theor}\ }\textbf {\bibinfo {volume} {48}},\
  \bibinfo {pages} {28FT01} (\bibinfo {year} {2015})}\BibitemShut {NoStop}%
\bibitem [{\citenamefont {Olshanii}(1998)}]{olshanii1998atomic}%
  \BibitemOpen
  \bibfield  {author} {\bibinfo {author} {\bibfnamefont {M.}~\bibnamefont
  {Olshanii}},\ }\href@noop {} {\bibfield  {journal} {\bibinfo  {journal} {Phy.
  Rev. Lett.}\ }\textbf {\bibinfo {volume} {81}},\ \bibinfo {pages} {938}
  (\bibinfo {year} {1998})}\BibitemShut {NoStop}%
\bibitem [{\citenamefont {Astrakharchik}\ \emph {et~al.}(2004)\citenamefont
  {Astrakharchik}, \citenamefont {Blume}, \citenamefont {Giorgini},\ and\
  \citenamefont {Granger}}]{astrakharchik2004quasi}%
  \BibitemOpen
  \bibfield  {author} {\bibinfo {author} {\bibfnamefont {G.}~\bibnamefont
  {Astrakharchik}}, \bibinfo {author} {\bibfnamefont {D.}~\bibnamefont
  {Blume}}, \bibinfo {author} {\bibfnamefont {S.}~\bibnamefont {Giorgini}}, \
  and\ \bibinfo {author} {\bibfnamefont {B.}~\bibnamefont {Granger}},\
  }\href@noop {} {\bibfield  {journal} {\bibinfo  {journal} {Phys. Rev. Lett.}\
  }\textbf {\bibinfo {volume} {92}},\ \bibinfo {pages} {030402} (\bibinfo
  {year} {2004})}\BibitemShut {NoStop}%
\bibitem [{\citenamefont {Chiofalo}\ \emph {et~al.}(2000)\citenamefont
  {Chiofalo}, \citenamefont {Succi},\ and\ \citenamefont
  {Tosi}}]{chiofalo2000ground}%
  \BibitemOpen
  \bibfield  {author} {\bibinfo {author} {\bibfnamefont {M.~L.}\ \bibnamefont
  {Chiofalo}}, \bibinfo {author} {\bibfnamefont {S.}~\bibnamefont {Succi}}, \
  and\ \bibinfo {author} {\bibfnamefont {M.}~\bibnamefont {Tosi}},\ }\href@noop
  {} {\bibfield  {journal} {\bibinfo  {journal} {Phys. Rev. E}\ }\textbf
  {\bibinfo {volume} {62}},\ \bibinfo {pages} {7438} (\bibinfo {year}
  {2000})}\BibitemShut {NoStop}%
\bibitem [{\citenamefont {Bao}\ and\ \citenamefont
  {Cai}(2012)}]{bao2012mathematical}%
  \BibitemOpen
  \bibfield  {author} {\bibinfo {author} {\bibfnamefont {W.}~\bibnamefont
  {Bao}}\ and\ \bibinfo {author} {\bibfnamefont {Y.}~\bibnamefont {Cai}},\
  }\href@noop {} {\bibfield  {journal} {\bibinfo  {journal} {arXiv preprint
  arXiv:1212.5341}\ } (\bibinfo {year} {2012})}\BibitemShut {NoStop}%
\bibitem [{\citenamefont {Pitaevskii}\ and\ \citenamefont
  {Stringari}(2003)}]{pitaevskii2003bose}%
  \BibitemOpen
  \bibfield  {author} {\bibinfo {author} {\bibfnamefont {L.}~\bibnamefont
  {Pitaevskii}}\ and\ \bibinfo {author} {\bibfnamefont {S.}~\bibnamefont
  {Stringari}},\ }\href@noop {} {\emph {\bibinfo {title} {Bose-Einstein
  Condensation}}}\ (\bibinfo  {publisher} {Oxford Science Publications},\
  \bibinfo {year} {2003})\BibitemShut {NoStop}%
\bibitem [{\citenamefont {Muryshev}\ \emph {et~al.}(1999)\citenamefont
  {Muryshev}, \citenamefont {van Linden van~den Heuvell},\ and\ \citenamefont
  {Shlyapnikov}}]{PhysRevA.60.R2665}%
  \BibitemOpen
  \bibfield  {author} {\bibinfo {author} {\bibfnamefont {A.~E.}\ \bibnamefont
  {Muryshev}}, \bibinfo {author} {\bibfnamefont {H.~B.}\ \bibnamefont {van
  Linden van~den Heuvell}}, \ and\ \bibinfo {author} {\bibfnamefont {G.~V.}\
  \bibnamefont {Shlyapnikov}},\ }\href {\doibase 10.1103/PhysRevA.60.R2665}
  {\bibfield  {journal} {\bibinfo  {journal} {Phys. Rev. A}\ }\textbf {\bibinfo
  {volume} {60}},\ \bibinfo {pages} {R2665} (\bibinfo {year}
  {1999})}\BibitemShut {NoStop}%
\bibitem [{\citenamefont {Akhmediev}\ and\ \citenamefont
  {Ankiewicz}(1993)}]{akhmediev1993first}%
  \BibitemOpen
  \bibfield  {author} {\bibinfo {author} {\bibfnamefont {N.}~\bibnamefont
  {Akhmediev}}\ and\ \bibinfo {author} {\bibfnamefont {A.}~\bibnamefont
  {Ankiewicz}},\ }\href@noop {} {\bibfield  {journal} {\bibinfo  {journal}
  {Phys. Rev. A}\ }\textbf {\bibinfo {volume} {47}},\ \bibinfo {pages} {3213}
  (\bibinfo {year} {1993})}\BibitemShut {NoStop}%
\end{thebibliography}%
\end{document}